\documentclass[letter,11pt]{article}
 \pdfoutput=1
 \usepackage{jheppub}
 
 \usepackage{graphicx}
 \usepackage{hyperref}
 \usepackage[utf8]{inputenc}
 \usepackage{amssymb}
\usepackage{multirow}
\usepackage{soul}
 
 \usepackage{booktabs}
\usepackage{mathrsfs}
\usepackage{epsfig}
\usepackage{graphicx}
\usepackage{subfigure}
\usepackage{dcolumn}
\usepackage{bm}
\usepackage{amsmath}
\usepackage{slashed}
\usepackage{multirow}
\usepackage{color}
\usepackage{dcolumn}
\usepackage{bm}
\usepackage{color}

\allowdisplaybreaks

 \newcommand{\lsim}{{\;\raise0.3ex\hbox{$<$\kern-0.75em\raise-1.1ex\hbox{$\sim$}}\;}}
\newcommand{\gsim}{{\;\raise0.3ex\hbox{$>$\kern-0.75em\raise-1.1ex\hbox{$\sim$}}\;}}
\newcommand{\beq}{\begin{equation}}
\newcommand{\eeq}{\end{equation}}
\newcommand{\bea}{\begin{eqnarray}}
\newcommand{\eea}{\end{eqnarray}}

\def\baa{\begin{array}}
\def\eaa{\end{array}}

\mathchardef\minus="002D

\preprint{ }

\title{Probing bino NLSP at lepton colliders}

\author{Junmou Chen$^1$,}
\emailAdd{chenjm@jnu.edu.cn}

\author{Chengcheng Han$^2$,}
\emailAdd{hanhch@mail.sysu.edu.cn}

\author{Jin Min Yang$^{3,4}$, }
\emailAdd{jmyang@itp.ac.cn}

\author{Mengchao Zhang$^1$}
\emailAdd{mczhang@jnu.edu.cn}

\vspace{0.5cm}

\affiliation{$^1$Department of Physics and Siyuan Laboratory, Jinan University, Guangzhou 510632, P.R. China}
\affiliation{$^2$School of Physics, Sun Yat-Sen University, Guangzhou 510275, P. R. China}
\affiliation{$^3$CAS Key Laboratory of Theoretical Physics, Institute of Theoretical
                Physics, Chinese Academy of Sciences, Beijing 100190, P. R. China}
\affiliation{$^4$School of Physical Sciences, University of Chinese Academy of Sciences,Beijing 100049, P. R. China}

\vspace{0.5cm}

\abstract{We consider a scenario where light bino is the next-to-lightest supersymmetric 
particle (NLSP) and gravitino/axino is the lightest superysmmetric particle (LSP). 
For a bino mass less than or around hundred GeV, it can be pair produced at the future lepton 
colliders through $t-$channel slepton exchange, subsequently decaying into 
a gravitino/axino plus a photon.   
We study the prospects to look for such binos at the future colliders and find that a bino
mass around 100 GeV can be probed at the $2\sigma$ ($5\sigma$) level for a slepton below 2 TeV (1.5 TeV) 
with a luminosity 3 $ab^{-1}$. For a bino mass around 10 GeV, a slepton mass less than 
4 TeV (3 TeV) can be probed at the $2\sigma$ ($5\sigma$) level, which is much beyond the reach of the LHC for direct slepton searches.
 }

\makeatletter
\def\@fpheader{\relax}
\makeatother

\date{2020.06.15}

\begin{document} 
\maketitle
\flushbottom
\newpage

\section{Introduction}
Despite of no experimental evidence, supersymmetry remains one of the most compelling 
scenarios beyond the standard model. It not only alleviates the fine-tuning of Higgs mass 
contrasting to the fundamental scale, but also predicts the unification of the gauge 
couplings and provides a viable dark matter candidate. Currently the large hadron 
colliders (LHC) already set very strong limits on the mass scale of SUSY partners. 
For example, the gluino and squarks should be beyond 1-2 TeV \cite{Aad:2020nyj, Aad:2020sgw} 
and the limits on electroweakinos and sleptons are around few hundreds GeV depending on the 
assumptions \cite{Aad:2019vvi, Aad:2019qnd}. For a degenerate spectrum such as higgsinos 
or winos, the strongest bounds are from the LEP and their masses should be larger than 
around 100 GeV \cite{LEP}. However, there remains a possibility that a bino could be as 
light as few tens of GeV. Given the design of a future lepton colliders \cite{CEPCStudyGroup:2018ghi, ILC, Abada:2019zxq}, it provides a good opportunity to look for such a light bino. 

If bino is the lightest supersymmetric particle (LSP) as well as the dark matter 
candidate, it usually has over-abundance for a light bino below 100 GeV. 
For a well-temperated bino-higgsino mixing dark matter, it is already excluded 
by dark matter direct searches for the light mass 
region \cite{Badziak:2017the,Han:2014sya,Abdughani:2017dqs}. 
Therefore it is reasonable to consider that a light bino might be the next-to-lightest 
supersymmetric particle (NLSP), while an additional particle is taken as the LSP. 
One of the natural LSP candidates is the gravitino, which could be much lighter than GeV 
for a low SUSY breaking scale such as in the gauge mediation models. Another possibility is that the LSP could be the axino, 
which is predicted by the breaking of Peccei–Quinn symmetry to solve the strong CP problem in the framework of 
supersymmetry \cite{Kim:1983dt, Barenboim:2014kka}. In both cases, the bino could probably 
decay into a photon and a LSP, providing a good signal at colliders. Such searches were 
already performed at the LHC where the gluino pair production sequentially decaying into 
neutralino and two quarks and the neutralino later decays into a photon and 
a gravitino \cite{ATLASCollaboration:2016wlb}.  Here we consider the bino pair production 
at lepton colliders through exchanging a t-channel slepton, and then the bino decays 
in the channel $\tilde{B} \rightarrow \tilde{G}/\tilde{a} + \gamma$
\footnote{There are also $\tilde{B} \rightarrow \tilde{G}/\tilde{a} + h/Z$ decay channels. However, for the bino mass region we are interested in these decay channels are suppressed by phase space, therefore we just ignore them and consider the branching ratio of $\tilde{B} \rightarrow \tilde{G}/\tilde{a} + \gamma$ is 100\%. }.
The typical signal is two photons plus large missing energy. Since the background 
is very clean at the lepton colliders, we expect such a signal could be detected 
for a selectron mass less than a few TeV and it would be much beyond the slepton 
searches at the LHC.

This paper is organized as follows: in section \ref{collider} we discuss the related 
collider analysis including the kinetic variables for cutting.  In section \ref{numerical} 
we present our numerical results and we draw our conclusion in section \ref{conclusion}.

\section{Collider analysis}
\label{collider}
\begin{figure*}[b!]
\centering
\includegraphics[width=1.0\textwidth]{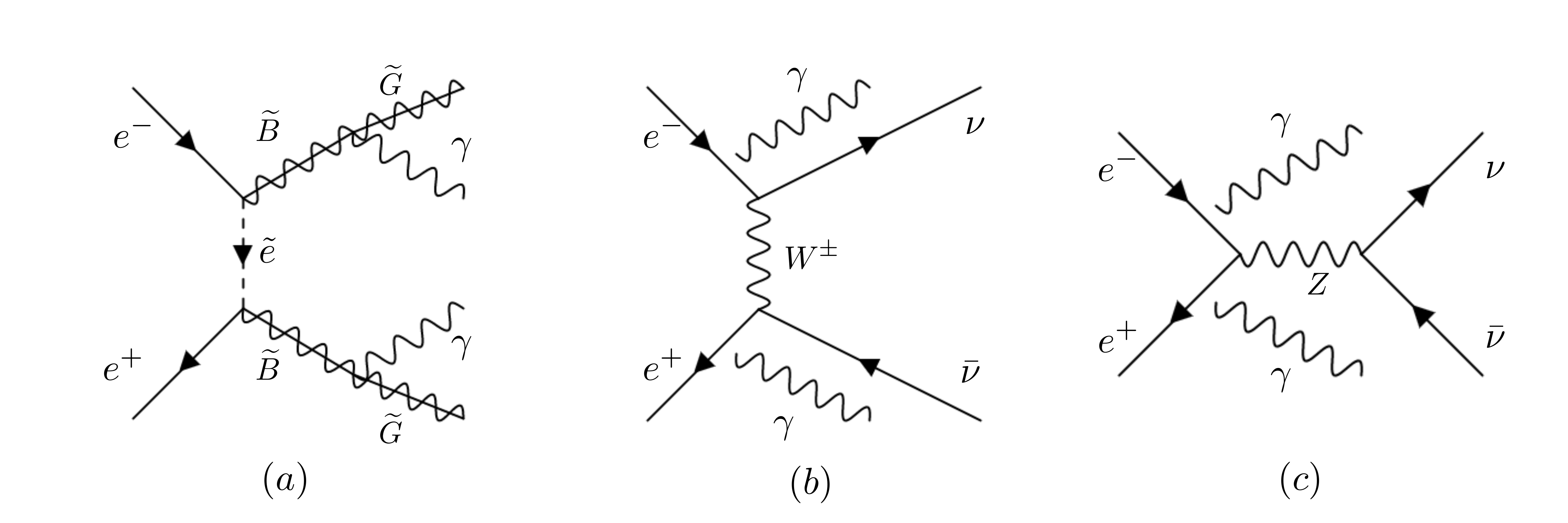}
\caption{(a) Feynman diagram of signal process; (b) Feynman diagrams of background process 
with t-channel $W$ boson and (c) Feynman diagrams of background process with s-channel $Z$ 
boson, where two photons can couple to any charged particles. }
\label{FD} 
\end{figure*}

In this section we analyse the collider signatures of the bino pair production at the 
lepton colliders. As a benchmark model, we study the scenario of gravitino as the LSP. 
The signal we consider is $e^+ e^- \to \tilde{B}\tilde{B} \to \gamma\gamma \tilde{G}\tilde{G}$.
The main SM background is $e^+ e^- \to \gamma\gamma \nu\bar{\nu}$, with $Z$ or $W^{\pm}$ boson 
being the internal particles. In Fig.~\ref{FD} we show the Feynman diagrams to illustrate 
our signal and background processes.

Both the signal and backgrounds are induced by electroweak interaction, 
but the signal is suppressed by the TeV-scale selectron and thus the production 
cross section is much smaller. 

In the following we calculate the cross section of bino pair production at an electron-positron 
collider with center-of-mass energy $\sqrt{s}$. 
For $e_R^-e^+_L\rightarrow \tilde{B}\tilde{B}$, where the bino production is mediated by 
$\tilde{e}_R$, the corresponding total cross section is 
\begin{eqnarray}
\sigma_{\text{tot}}^R&=&\frac{g_1^4Y_{e_R}^4}{8\pi s^2} \left[ 4E|\vec{k}| -
\left(2m_{\tilde{e}_R}^2- 2m_{\tilde{B}}^2\frac{2(m_{\tilde{e}_R}^2-m_{\tilde{B}}^2)}{s+2(m_{e_R}^2-m_{\tilde{B}}^2)}\right) 
\log\frac{(2E(E+|\vec{k}|)+ m_{\tilde{e}_R}^2-m_{\tilde{B}}^2)^2}{4E^2m_{\tilde{e}_R}^2+ (m_{\tilde{e}_R}^2-m_{\tilde{B}}^2)^2} 
\right]\nonumber \\
&&+ \frac{g_1^4Y_{e_R}^4}{8\pi s^2}(m_{\tilde{e}_R}^2-m_{\tilde{B}}^2)^2 \frac{4E(E+|\vec{k}|)}{4E^2m_{\tilde{e}_R}^2
+ (m_{\tilde{e}_R}^2-m_{\tilde{B}}^2)^2} .
\end{eqnarray}
For $e_L^-e^+_R\rightarrow \tilde{B}\tilde{B}$ process which is mediated by $\tilde{e}_L$, 
the corresponding total cross section is
\begin{eqnarray}
\sigma_{\text{tot}}^L&=&\frac{g_1^4Y_{e_L}^4}{8\pi s^2} \left[ 4E|\vec{k}| -
\left(2m_{\tilde{e}_L}^2- 2m_{\tilde{B}}^2\frac{2(m_{\tilde{e}_L}^2-m_{\tilde{B}}^2)}{s+2(m_{e_L}^2-m_{\tilde{B}}^2)}\right) 
\log\frac{(2E(E+|\vec{k}|)+ m_{\tilde{e}_L}^2-m_{\tilde{B}}^2)^2}{4E^2m_{\tilde{e}_L}^2+ (m_{\tilde{e}_L}^2-m_{\tilde{B}}^2)^2} 
\right]\nonumber \\
&&+ \frac{g_1^4Y_{e_L}^4}{8\pi s^2}(m_{\tilde{e}_L}^2-m_{\tilde{B}}^2)^2 \frac{4E(E+|\vec{k}|)}{4E^2m_{\tilde{e}_L}^2
+ (m_{\tilde{e}_L}^2-m_{\tilde{B}}^2)^2} .
\end{eqnarray}
In the above expressions, $k^{\mu}=(E,\vec{k})$ is the 4-momentum of 
one of the binos and $s= 4E^2$ is the central energy square.  We have set $m_e=0$.
If $\tilde{e}_R$ and $\tilde{e}_L$ have the same masses, then the ratio between 
$\sigma_{\text{tot}}^R$ and $\sigma_{\text{tot}}^L$ is
\begin{equation}
\frac{\sigma_{\text{tot}}^L}{\sigma_{\text{tot}}^R} = \frac{Y^4_{e_L}}{Y^4_{e_R}}= \frac{1}{16}  .
\end{equation}
For simplicity, in the following we assume the masses of $\tilde{e}_R$ and $\tilde{e}_L$ 
are equal and define $m_{\tilde{e}} \equiv m_{\tilde{e}_R} = m_{\tilde{e}_L} $.
Then the total cross section can be simplified to the following formula if selectrons 
are heavy enough:
\begin{equation}
\sigma_{\text{tot}} (e^+ e^- \to \tilde{B}\tilde{B}) \approx \frac{g_1^4}{48\pi} 
\left( {Y^4_{e_L}}+{Y^4_{e_R}} \right) \frac{s}{m^4_{\tilde{e}}} 
\left[1-\left(\frac{m_{\tilde{B}}}{E} \right)^2\right]^{3/2} .
\label{xs}
\end{equation}

\begin{figure*}[b!]
\centering
\includegraphics[width=0.8\textwidth]{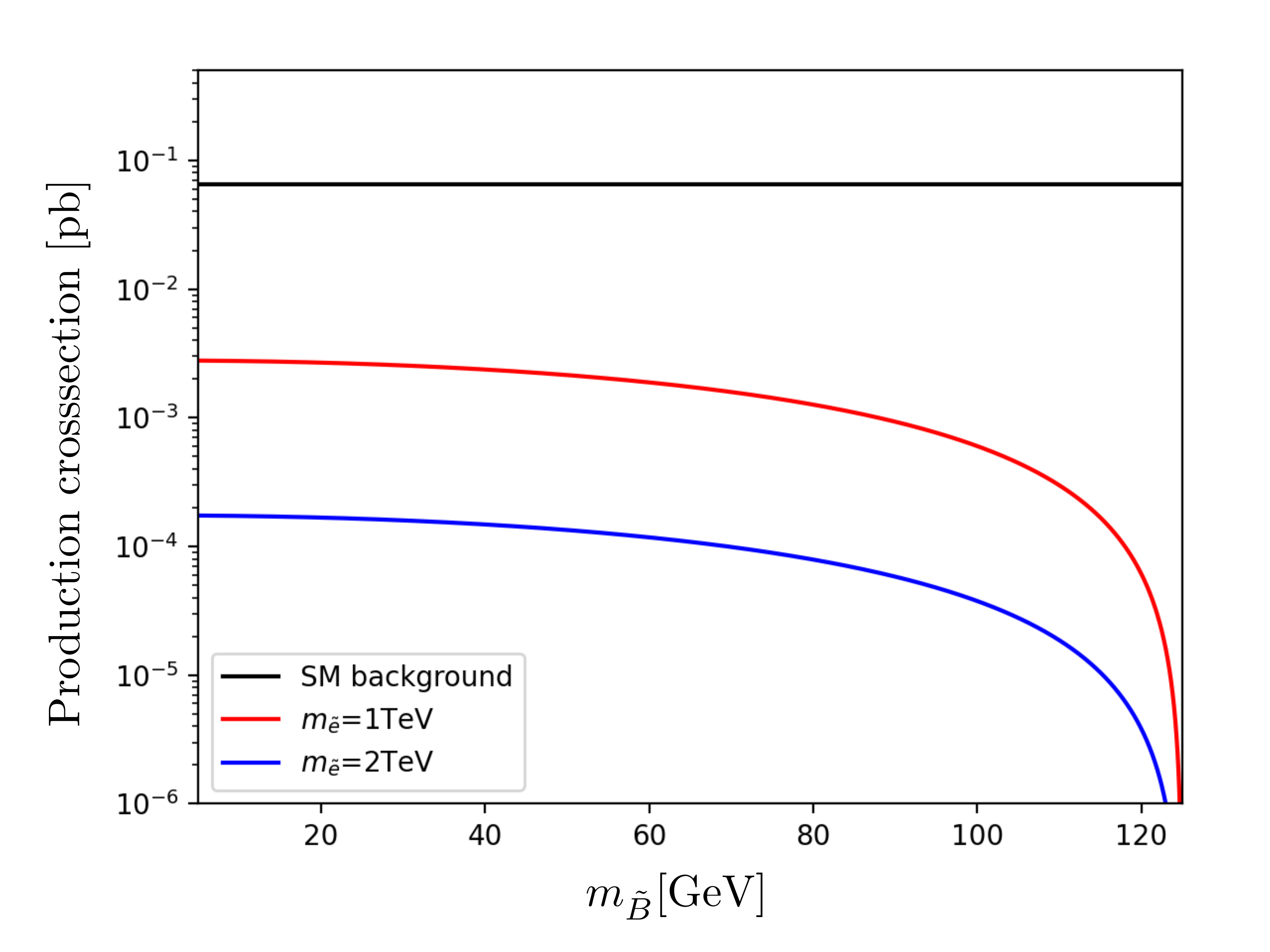}
\caption{The black line is the cross section of 
$\sigma(e^+ e^- \to \gamma\gamma \nu\bar{\nu})$ in the SM, with photon $p_{\text{T}} > 10$ GeV.
The red and blue curves are the cross section of $\sigma(e^+ e^- \to \tilde{B}\tilde{B})$ 
with $m_{\tilde{e}}$ being 1 TeV and 2 TeV, respectively.
The center-of-mass energy is fixed to 250 GeV.}
\label{Xs} 
\end{figure*}

In Fig.~\ref{Xs} we show the cross sections of $e^+ e^- \to \tilde{B}\tilde{B}$ 
and the SM background $e^+ e^- \to \gamma\gamma \nu\bar{\nu}$ for $\sqrt{s}$=250 GeV. 
Here we use {\tt MadGrpaph5}~\cite{Alwall:2011uj} to calculate the cross section of $e^+ e^- \to \gamma\gamma \nu\bar{\nu}$.
Due to the soft and collinear divergence when photons are radiated from the initial state electron and positron, 
we require the transverse momentums of photons in the final state to be larger than 10 GeV. 
It is clear that a TeV-scale $m_{\tilde{e}}$ highly suppresses the cross section of the signal.
To improve the direct search sensitivity, additional cuts are needed.  
The final state of the signal and background processes are extremely simple and clear, 
only two photons and missing momentum. What we can directly measure are the 
4-momentums of the two final state photons:
\begin{eqnarray}
p^{\gamma 1} &=& \left( E^{\gamma 1} \ , \ p_x^{\gamma 1} \ , \ p_y^{\gamma 1} \ , \ p_z^{\gamma 1}  \right) \\
p^{\gamma 2} &=& \left( E^{\gamma 2} \ , \ p_x^{\gamma 2} \ , \ p_y^{\gamma 2} \ , \ p_z^{\gamma 2}  \right)
\end{eqnarray}
These 8 components will be the inputs of our analysis and all the kinetic variables 
will be constructed from them. 
For Monte Carlo simulation, we use {\tt MadGrpaph5}~\cite{Alwall:2011uj} to generate the 
parton-level signal and background events. 
QED shower, which can happen when the energy of charged particle is much larger than its mass, 
is performed by {\tt PYTHIA}~\cite{Sjostrand:2014zea}.
Considering the detector effect, we perform a Gaussian smearing on the photon 
energy measurement with uncertainty 3\%~\cite{Shen:2019yhf}. 

\subsection{Variable-1: recoiled mass}
The first kinetic variable we will use is the recoiled mass, which has been proven 
to be a useful variable in Higgs measurements at lepton colliders~\cite{Gu:2017del}.  
Unlike the hadron collider with undermined initial-state parton 4-momentum, the 4-momentums of 
the initial-state electron and positron are known. For a lepton collider with 
center-of-mass energy square $s$, the signal contains two photons in the final state. 
Thus from the 4-momentum conservation we have
\begin{eqnarray}
(\sqrt{s},0,0,0) = p^{\gamma 1} + p^{\gamma 2} + p\!\!\!/
\end{eqnarray}
By knowing $p^{\gamma 1}$ and $p^{\gamma 2}$, we can construct the recoiled mass as
\begin{eqnarray}
m_{\text{rec}} = \sqrt{   ((\sqrt{s},0,0,0) - p^{\gamma 1} - p^{\gamma 2})^2   } 
\end{eqnarray}
We expect this recoiled mass to peak around the $Z$ boson mass for the SM background process 
because the SM background is dominated by the s-channel $Z$ boson process. 
And for our signal process, the distribution of $m_{\text{rec}}$ should be much more flat 
because the missing momentum in signal process comes from two gravitinos of two branches. 

\begin{figure*}[tbp!]
\centering
\includegraphics[width=0.8\textwidth]{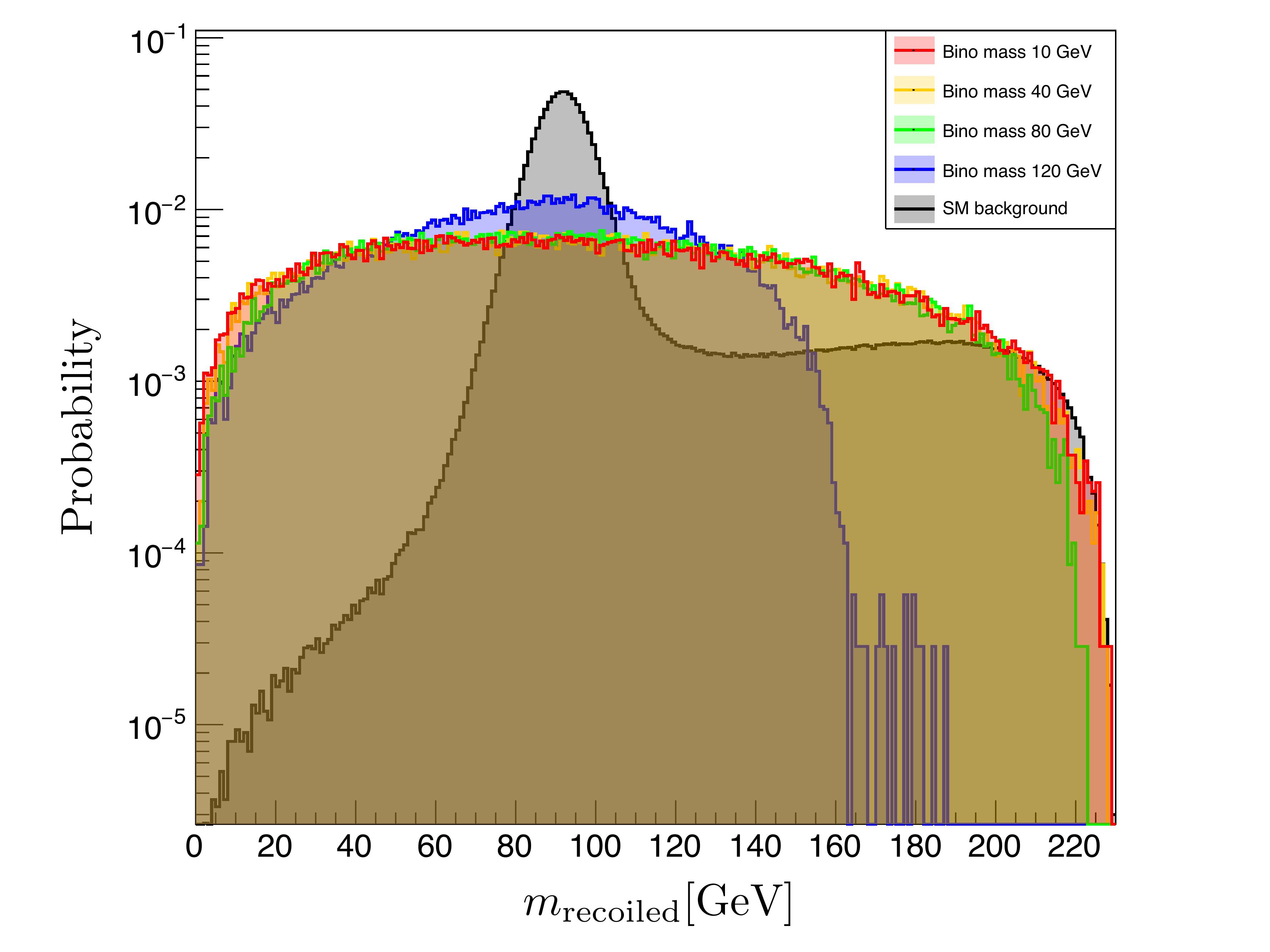}
\caption{Recoiled mass distributions for the signal and SM background processes with different bino masses.}
\label{recoil_m} 
\end{figure*}

In Fig.~\ref{recoil_m} we present the recoiled mass distribution for the SM background and signal process 
with different bino masses.
As expected, the SM distribution peaks around 90 GeV, which comes from the s-channel $Z$ boson. 
The recoiled mass from t-channel $W$ boson tends to have a larger value, and the distribution 
extends smoothly to the kinetically allowed region. 
On the contrary, the recoiled mass of signal process is basically evenly distributed.  
Therefore, the background can be hugely suppressed by requiring an upper limit of $m_{\text{recoiled}}$.

\subsection{Variable-2: reconstructed mass}\
Since there are two decay branches with invisible particles in the final states, 
the mass of bino can not be easily reconstructed. 
At hadron collider, one can use the variables like MT2~\cite{Burns:2008va,Cho:2007qv,Matchev:2009ad,Konar:2009wn} to discriminate signal from SM background.
At lepton collider, momentums in longitudinal direction can also be measured, and thus the kinematics are different with hadron collider.
Here we propose a kinetic variable which can reflect the bino mass through its distribution.

In addition to the 8 known values (4-momentums of two photons), there are 8 unknown values 
which come from invisible gravitinos:
\begin{eqnarray}
p^{\tilde{G} 1} &=& \left( E^{\tilde{G} 1} \ , \ p_x^{\tilde{G} 1} \ , \ p_y^{\tilde{G} 1} \ , \ p_z^{\tilde{G} 1}  \right) \\
p^{\tilde{G} 2} &=& \left( E^{\tilde{G} 2} \ , \ p_x^{\tilde{G} 2} \ , \ p_y^{\tilde{G} 2} \ , \ p_z^{\tilde{G} 2}  \right)
\end{eqnarray}
The missing momentum is composed by these two invisible momentums. For a lepton collider, we know the 4-momentums 
of initial states. Thus the missing momentum can be calculated from the momentums of two photons, 
and we obtain 4 constraint equations for 8 unknown values:
\begin{eqnarray}
E^{\tilde{G} 1} + E^{\tilde{G} 2} &=& \sqrt{s} -  E^{\gamma 1} -E^{\gamma 2} \\
p_x^{\tilde{G} 1} + p_x^{\tilde{G} 2} &=& - p_x^{\gamma 1} -p_x^{\gamma 2} \\
p_y^{\tilde{G} 1} + p_y^{\tilde{G} 2} &=&  -  p_y^{\gamma 1} -p_y^{\gamma 2} \\
p_z^{\tilde{G} 1} + p_z^{\tilde{G} 2} &=&  -  p_z^{\gamma 1} -p_z^{\gamma 2} 
\end{eqnarray}
with $\sqrt{s}$ being the center-of-mass energy of the $e^+ e^-$ collider. 
By using these 4 constraint conditions, the number of unknown values decrease from 8 to 4. 
We can further limit those unknown values by using the mass of gravitino $m_{\tilde{G}}$, 
which is smaller than GeV.
Such a tiny mass can be treated as zero for our collider kinematic analysis. 
Then the on-shell conditions give two more constraint equations:
\begin{eqnarray}
\left( E^{\tilde{G} 1} \right)^2  &=& \left( p_x^{\tilde{G} 1} \right)^2 
+ \left( p_y^{\tilde{G} 1} \right)^2 + \left( p_z^{\tilde{G} 1} \right)^2   \\
\left( E^{\tilde{G} 2} \right)^2  &=& \left( p_x^{\tilde{G} 2} \right)^2 
+ \left( p_y^{\tilde{G} 2} \right)^2 + \left( p_z^{\tilde{G} 2} \right)^2 
\end{eqnarray}
and the number of unknown values decrease from 4 to 2. 
And we can re-express the 4-momentums of ${\tilde{G} 1}$ and ${\tilde{G} 2}$ as
\begin{eqnarray}
p^{\tilde{G} 1} &=& E^{\tilde{G} 1} (1 \ , \ \vec{e}_1 ) \\
p^{\tilde{G} 2} &=& E^{\tilde{G} 2}  (1 \ , \ \vec{e}_2 )
\end{eqnarray}
Here we use the unit 3-vectors $\vec{e}_1$ and $\vec{e}_2$ to represent the moving directions 
of ${\tilde{G} 1}$ and ${\tilde{G} 2}$, respectively.

\begin{figure*}[t!]
\centering
\includegraphics[width=0.8\textwidth]{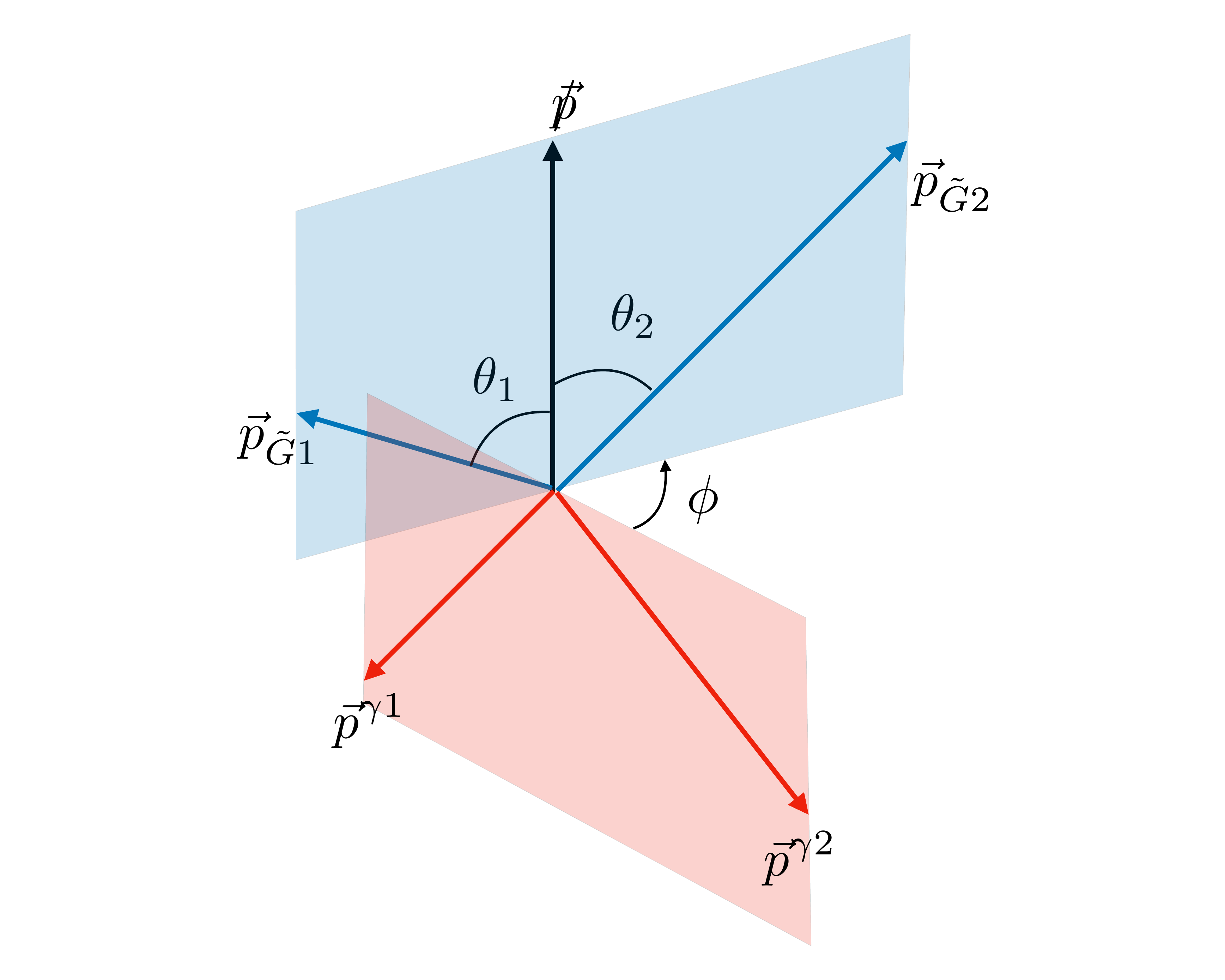}
\caption{Geometric relations between $\vec{p}_{\tilde{G} 1}$, $\vec{p}_{\tilde{G} 2}$ and the 3 measured momentums. 
$\vec{p}_{\tilde{G} 1}$, $\vec{p}_{\tilde{G} 2}$ and ${\vec{p}}\!\!\!/$ are in a plane, 
while $\vec{p}^{\gamma 1}$, $\vec{p}^{\gamma 2}$ and ${\vec{p}}\!\!\!/$ are in another plane. 
The angle $\phi$ between the two planes can be fixed by the measured values and kinetic relations.}
\label{vector} 
\end{figure*}

The last advantage of a lepton collider that we are going to use is that the two binos must be generated 
with the same speed. Thus the energies of these two binos must be equal. 
Combining with energy conservation, we obtain two linear equations that can be used to 
fix $E^{\tilde{G} 1}$ and $E^{\tilde{G} 2}$:
\begin{eqnarray}
\left\{
\begin{array}{rl}
E^{\tilde{G} 1} + E^{\tilde{G} 2} &= \sqrt{s} -  E^{\gamma 1} -E^{\gamma 2}\\
E^{\tilde{G} 1} + E^{\gamma 1} &= E^{\tilde{G} 2} + E^{\gamma 2}
\end{array}
\right. 
\Rightarrow 
\left\{
\begin{array}{rl}
E^{\tilde{G} 1}  &= \frac{\sqrt{s}}{2} -  E^{\gamma 1}  \\
E^{\tilde{G} 2}  &= \frac{\sqrt{s}}{2} -  E^{\gamma 2} 
\end{array}
\right.
\label{condition1}
\end{eqnarray}
Now we only have one unknown value in this kinematic system. 
In the following we explain what is the last ambiguity. 
We label the 3-momentums of the two binos and total missing momentum as 
$\vec{p}_{\tilde{G} 1}$, $\vec{p}_{\tilde{G} 2}$ and ${\vec{p}}\!\!\!/$:
\begin{eqnarray}
\vec{p}_{\tilde{G} 1} = E^{\tilde{G} 1} \vec{e}_1   \ ,~~ 
\vec{p}_{\tilde{G} 2} = E^{\tilde{G} 2} \vec{e}_2   \ , ~~
{\vec{p}}\!\!\!/ =   \left(-p_x^{\gamma 1}-p_x^{\gamma 2}  \ , -p_y^{\gamma 1}-p_y^{\gamma 2} \ ,-p_z^{\gamma 1}-p_z^{\gamma 2} \right)
\end{eqnarray}
We already knew ${\vec{p}}\!\!\!/$ and the lengthes of $\vec{p}_{\tilde{G} 1}$ and $\vec{p}_{\tilde{G} 2}$.
Because $\vec{p}_{\tilde{G} 1} + \vec{p}_{\tilde{G} 2} = {\vec{p}}\!\!\!/$, so these 3 vectors must be in the same plane. 
On the other hand, the total 3-momentum of final states should be zero, 
so $\vec{p}^{\gamma 1}$, $\vec{p}^{\gamma 2}$, and ${\vec{p}}\!\!\!/$ are in another plane.
For illustration, we show their geometry in Fig~\ref{vector}. 
Fixing the lengthes of $\vec{p}_{\tilde{G} 1}$ and $\vec{p}_{\tilde{G} 2}$, 
we can determine the angle between them and ${\vec{p}}\!\!\!/$ (noted as $\theta_1$ and $\theta_2$ in  Fig~\ref{vector}).
But the angle between the two different planes (noted as $\phi$ in Fig~\ref{vector}), 
can not be determined by input values. 
So the value of $\phi$ is the last one that we can not fix. 

\begin{figure*}[t!]
\centering
\includegraphics[width=0.8\textwidth]{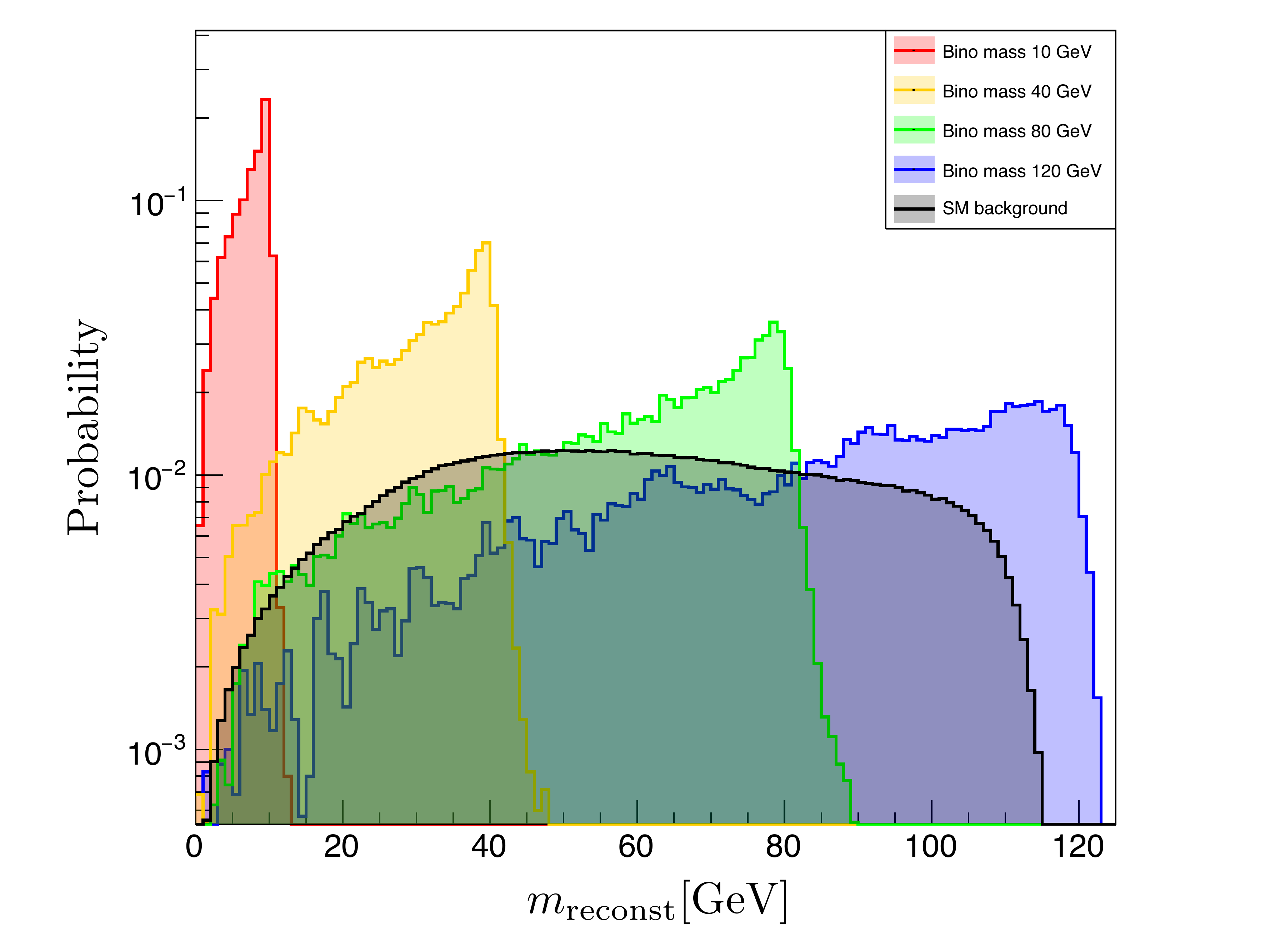}
\caption{Reconstructed mass distributions for the signal and SM background processes with different bino masses.}
\label{reconst_m} 
\end{figure*}

The value range of $\phi$ is 0 to $2\pi$. 
We can calculate the trial mass of bino (noted as $m^*_{\tilde{B}1}$ and $m^*_{\tilde{B}2}$) at each branch 
as a function of $\phi$: 
\begin{eqnarray}
m^*_{\tilde{B}1}(\phi) &=& \sqrt{ (E^{\tilde{G} 1} + E^{\gamma 1}  )^2 - (\vec{p}_{\tilde{G} 1} + \vec{p}^{\gamma 1} )^2   } \\\nonumber
&=&  \sqrt{ 2 E^{\tilde{G} 1} E^{\gamma 1}  - 2 \vec{p}_{\tilde{G} 1} \cdot \vec{p}^{\gamma 1}}   \\
m^*_{\tilde{B}2}(\phi) &=& \sqrt{ (E^{\tilde{G} 2} + E^{\gamma 2}  )^2 - (\vec{p}_{\tilde{G} 2} + \vec{p}^{\gamma 2} )^2   } \\\nonumber
&=&  \sqrt{ 2 E^{\tilde{G} 2} E^{\gamma 2}  - 2 \vec{p}_{\tilde{G} 2} \cdot \vec{p}^{\gamma 2}}
\end{eqnarray}
where $\vec{p}_{\tilde{G} 1}$ and $\vec{p}_{\tilde{G} 2}$ are functions of $\phi$. 
Furthermore, $m^*_{\tilde{B}1}(\phi) = m^*_{\tilde{B}2}(\phi)$ holds for any value of $\phi$ 
because this condition has been used in Eq.~\ref{condition1}.
In the following we denote both $m^*_{\tilde{B}1}(\phi)$ and $m^*_{\tilde{B}2}(\phi)$ indiscriminately 
as $m^*_{\tilde{B}}(\phi)$.
$m^*_{\tilde{B}}(\phi)$ can be minimized with a certain value of $\phi$. 
This minimum value is equal to the real bino mass only when all the momentums in Fig~\ref{vector} are in a same plane, otherwise the minimum value of $m^*_{\tilde{B}}(\phi)$ is smaller than the real bino mass. 
Thus we define the \textit{reconstructed mass} as the minimum value of  $m^*_{\tilde{B}}(\phi)$:
\begin{eqnarray}
m_{\text{reconst}} = \min \limits_{\phi \in \left (0, 2\pi \right ) } {m^*_{\tilde{B}}(\phi)}
\end{eqnarray}
For the signal process, $m_{\text{reconst}}$ can not exceed the real bino mass.
In Fig.~\ref{reconst_m} we present the distributions of $m_{\text{reconst}}$ for the SM background 
and signal process with different bino masses.
It can be seen that for the signal process, $m_{\text{reconst}}$ basically distribute in the region that smaller than the real bino mass. 
The events that excess the bino mass limit come from the uncertainty in the photon energy measurement.
The distribution of $m_{\text{reconst}}$ peaks when it is close to the real bino mass.
On the other hand the \textit{reconstructed mass} for the background process is basically evenly distributed.  
Thus a simple mass window cut on $m_{\text{reconst}}$ can help us to enhance the search sensitivity. 

\begin{figure*}[b!]
\centering
\includegraphics[width=0.8\textwidth]{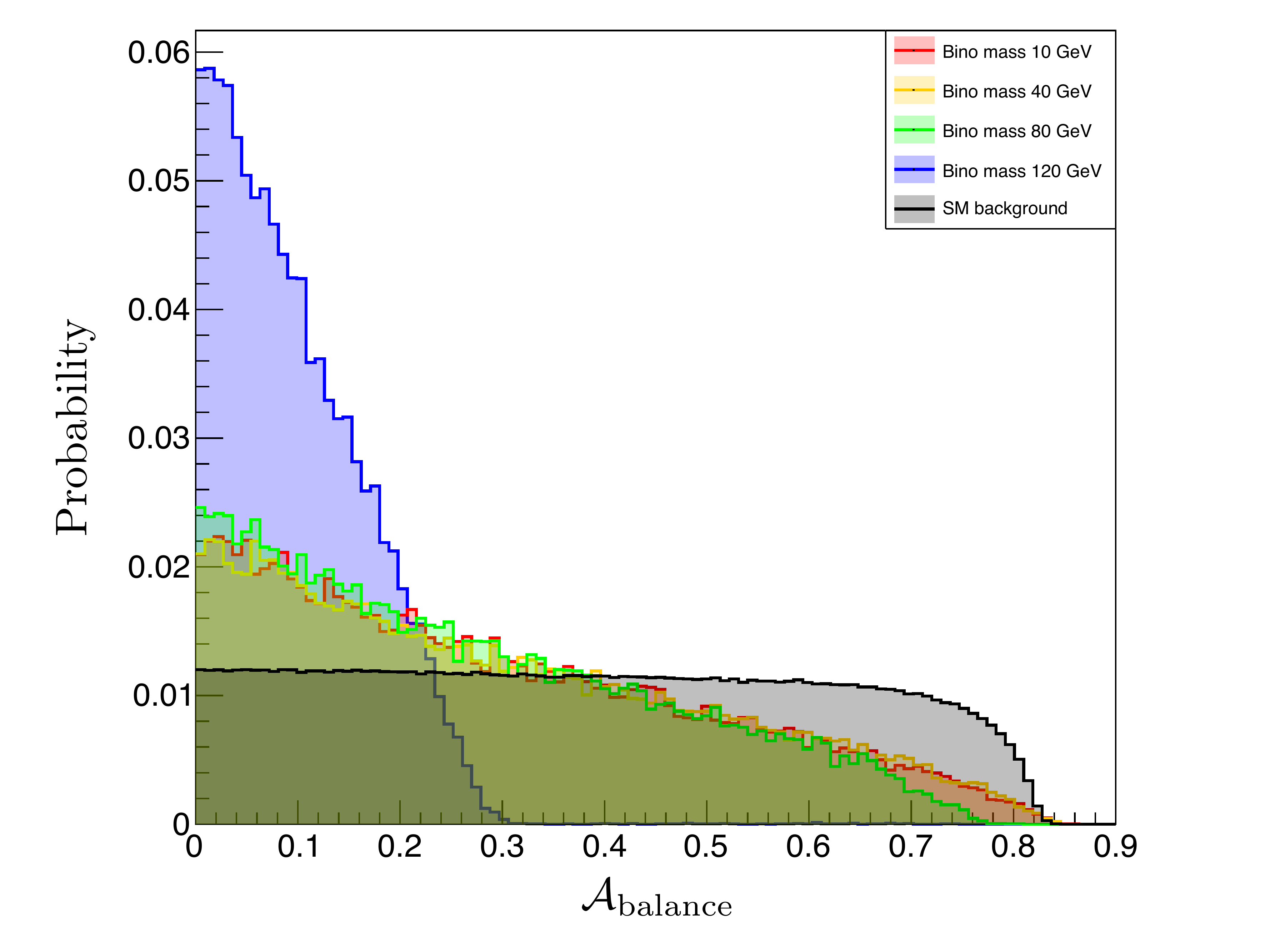}
\caption{Energy balance distributions for the signal and SM background processes with different bino masses.}
\label{balance} 
\end{figure*}

\subsection{Variable-3: energy balance}

To improve the sensitivity in the heavy bino region, 
we introduce another variable \textit{energy balance} which is defined as
\begin{eqnarray}
\mathcal{A}_{\text{balance}} = \frac{|E^{\gamma 1}-E^{\gamma 2}|}{|E^{\gamma 1}+E^{\gamma 2}|}
\end{eqnarray}
This variable can be very useful in signal and background discrimination when bino is heavy. 
In this case, the bino is not so boosted and the energy of each final-state photon is around 
the half of bino mass. 
Then the value of $\mathcal{A}_{\text{balance}}$ will be quite small when bino is heavy. 
On the contrary, if bino is light, the energy of each final-state photon is quite random 
(depending on the decay direction) and the distribution of $\mathcal{A}_{\text{balance}}$ 
will be more uniform. For the background process, the distribution should also be flat. 
In Fig.~\ref{balance} we present the distributions of $\mathcal{A}_{\text{balance}}$ 
for the SM background and signal processes with different bino masses.

\section{Numerical results}
\label{numerical}
In this section we use the three variables defined in the preceding section 
to perform cuts for our signal and backgrounds. 
It should be noted that the final state kinetic distributions are almost 
not relevant to the mass of selectron $m_{\tilde{e}}$ which only 
rescales the total cross section of the signal process roughly by a factor $1/m^4_{\tilde{e}}$.
To illustrate the performance of the cut-flow, we will fix $m_{\tilde{e}}$ to 1 TeV 
and vary $m_{\tilde{B}} = $ 10 GeV, 40 GeV, 80 GeV, 120 GeV as our benchmark points.  
The final results will be given in the end of this section. 
For future lepton colliders, we focus on a center-of-mass energy 
of 250 GeV and an integral luminosity of 3 ab$^{-1}$. 

The following cuts are imposed to enhance the search sensitivity:
\begin{itemize}
\item Recoiled mass $m_{\text{recoiled}}$ is required to be smaller than 40 GeV. 
The reason for this requirement can be seen from Fig.~\ref{recoil_m}. 
It is clear that the distribution of $m_{\text{recoiled}}$ for the background process 
rapidly decrease in the low $m_{\text{recoiled}}$ region. Thus removing the event 
with large $m_{\text{recoiled}}$ will suppress the background hugely.  

\item Energy balance $\mathcal{A}_{\text{balance}}$ is required to be smaller than 0.3. 
As can be seen from Fig.~\ref{balance}, the requirement helps to enhance the search 
sensitivity when bino is heavy. 

\item Mass window cut is performed on the reconstructed mass $m_{\text{reconst}}$:
\begin{eqnarray}
 m_{\tilde{B}} - \text{10 GeV}  \leqslant  m_{\text{reconst}} \leqslant m_{\tilde{B}} + \text{3 GeV}
\end{eqnarray}
This mass window cut depends on the bino mass. 
The upper and lower limits of mass window is chosen to optimize the search sensitivity. 
\end{itemize}

In Table ~\ref{cutflow} we present the cut flows for the SM background and our benchmark points. 
It shows that the $m_{\text{recoiled}}<$ 40 GeV cut reduce the SM background by three orders 
of magnitude, but only reduce the signal by one order of magnitude. The cuts of energy balance 
and mass window 
furtherly reduce the SM background by about one or two orders of magnitude, without hurting the 
signal too much.  We use $\mathcal{S}$ and $\mathcal{B}$ to represent the number of signal and 
background events after the cut flow, respectively. 
Then the statistical significance is evaluated by Poisson formula
\begin{eqnarray}\label{ss}
\text{Significance} =\sqrt{2\left[(\mathcal{S}+\mathcal{B})
         {\rm ln}(1+\frac{\mathcal{S}}{\mathcal{B}})-\mathcal{S}\right]}
\end{eqnarray}
For observation or exclusion, we require $\mathcal{S}/\mathcal{B}$ should be larger than 10\%  
and the significance should be larger than 5 or 2. 
In the last two rows of Table~\ref{cutflow} we show $\mathcal{S}/\mathcal{B}$ and the 
significance for each benchmark point. 
Therefore, for all 4 benchmark points 
the statistical significances are much larger than 5$\sigma$ for a slepton mass around 1 TeV. 

\begin{table}[htb]
\begin{center}\begin{tabular}{|c|c|c|c|c|c|}
\hline  & BKG & $m_{\tilde{B}}$=10 GeV & $m_{\tilde{B}}$=40 GeV & $m_{\tilde{B}}$=80 GeV & $m_{\tilde{B}}$=120 GeV \\
\hline photon $p_{\text{T}}>10$ GeV & 194,100 & 6,193 & 5,539 & 3,458 & 173 \\
\hline $m_{\text{recoiled}}<$  40 GeV & 147 & 912 & 785 & 439 & 18.6 \\
\hline $\mathcal{A}_{\text{balance}}<$ 0.3 & 40.4 & 661 & 553 & 317 & 18.6 \\
\hline $m_{\text{reconst}} \in [0,13]$ GeV & 1.1 & 651 & - & - & - \\
\hline $m_{\text{reconst}} \in [30,43]$ GeV & 2.8 & - & 269 & - & - \\
\hline $m_{\text{reconst}} \in [70,83]$ GeV & 3.5 & - & - & 96.8 & - \\
\hline $m_{\text{reconst}} \in [110,123]$ GeV  & 13.1 & - & - & - & 14.8 \\
\hline $\mathcal{S}/\mathcal{B}$ & - & 633.2 & 96.6 & 27.7 & 1.13 \\
\hline Significance & - & 83.8 & 44.1 & 21.9 & 3.5 \\
\hline \end{tabular} 
\caption{The cut flow for the SM background and 4 benchmark points of the signal. 
For the lepton collider we set 
the center-of-mass energy to be 250 GeV and the integral luminosity to be 3 ab$^{-1}$. 
For different benchmark points, we perform different mass window cuts on $m_{\text{reconst}}$.}
\end{center}
\label{cutflow}
\end{table}

\begin{figure*}[tb]
\centering
\includegraphics[width=1.0\textwidth]{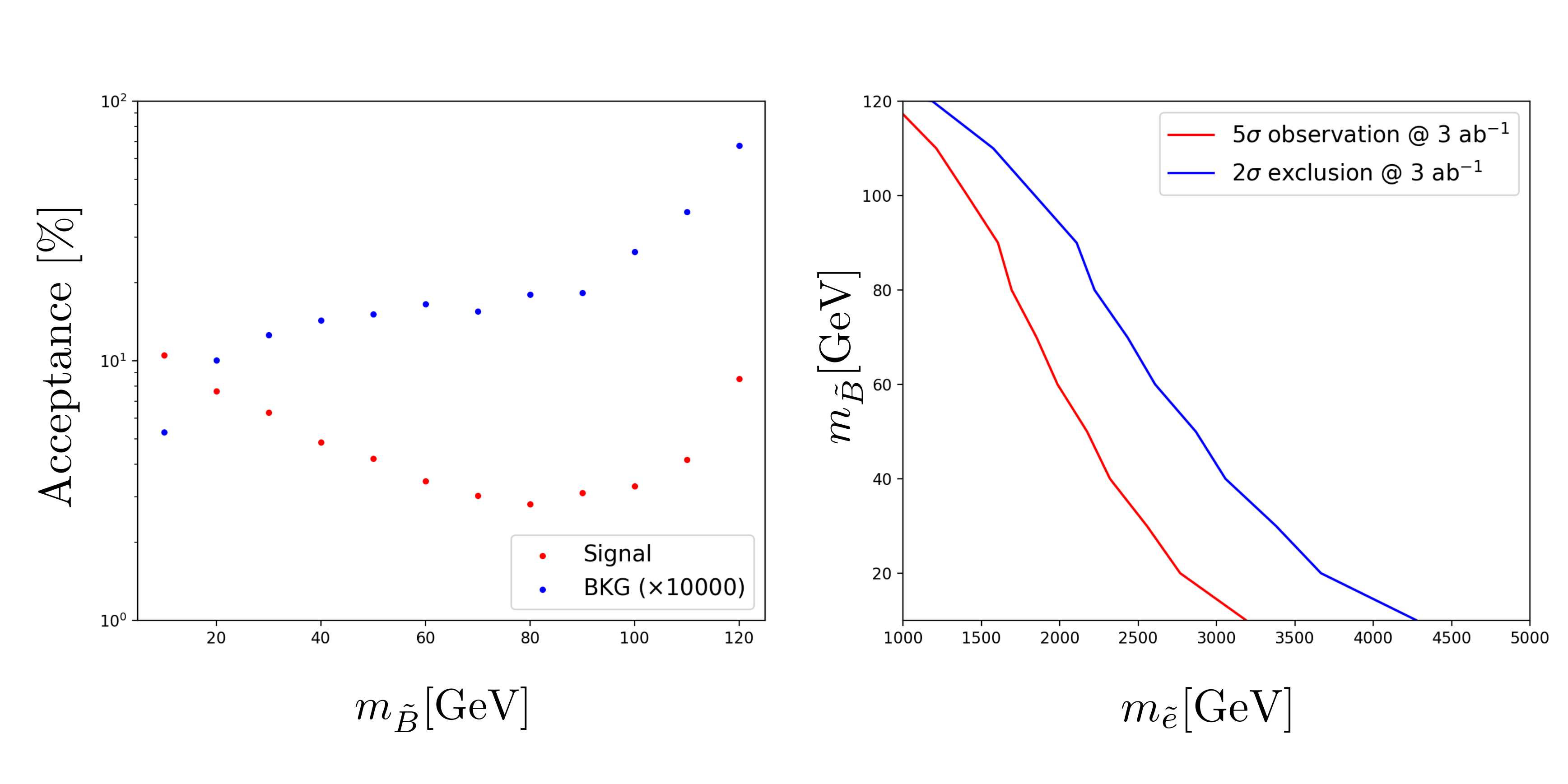}
\vspace*{-1.0cm}
\caption{\textit{Left}: acceptance of signal and background processes as functions of $m_{\tilde{B}}$. 
Here the acceptance of background process has been multiplied by 10,000. 
\textit{Right}: 2$\sigma$ exclusion and 5$\sigma$ observation limits on the $m_{\tilde{e}} - m_{\tilde{B}}$ plane 
at a future lepton collider running with an integral luminosity 3 ab$^{-1}$ and center-of-mass energy 250 GeV. 
Regions below the curves are observable or excluded.}
\label{limits} 
\end{figure*}

At the end, we can obtain the 2$\sigma$ exclusion limit (5$\sigma$ observation limit) 
on $m_{\tilde{e}} - m_{\tilde{B}}$ plane by requiring $\mathcal{S}/\mathcal{B}>0.1$ and 
Significance$> 2$ (Significance$> 5$). For each $m_{\tilde{B}}$ we obtain the signal acceptance 
by performing our cut-flow. The signal acceptance as a function of $m_{\tilde{B}}$ is given in 
Fig.~\ref{limits} (\textit{left}). We also present the acceptance of SM background as function of $m_{\tilde{B}}$ 
in Fig.~\ref{limits} (\textit{left})\footnote{Mass window cut depends on the value of $m_{\tilde{B}}$.}. 
By using these acceptance information and the cross section given by Eq.~\ref{xs}, we can easily 
calculate the significance for each spectrum $(m_{\tilde{e}}, m_{\tilde{B}})$, and thus find out the 
2$\sigma$ exclusion and 5$\sigma$ observation limits on the $m_{\tilde{e}} - m_{\tilde{B}}$ plane.
We show our final results in Fig.~\ref{limits} (\textit{right}).  It tells that a future 
lepton collider with an integral luminosity 3 ab$^{-1}$ and center-of-mass energy 250 GeV can 
exclude a bino below about 10 GeV for selectron less than 6 TeV, or a bino below about 100 GeV  
for a selectron less than 3 TeV, which is much beyond the reach of the LHC for direct slepton searches. 

\section{Conclusion}
\label{conclusion}
The bino is the only supersymmetric partner whose mass could be less than a hundred GeV. 
The future lepton colliders could provide a good opportunity to probe such a light particle. 
We considered a scenario where a light bino is the next-to-lightest superysmmetric particle (NLSP) 
and the gravition/axino is the lightest superysmmetric particle (LSP). Then the bino can probably 
decay into a photon plus the LSP. We studied the bino pair production at the future colliders and 
found that a bino mass around 100 GeV can be probed at the $2\sigma$ ($5\sigma$) level for a slepton below 2 TeV (1.5 TeV) 
with a luminosity 3 $ab^{-1}$. For a bino mass around 10 GeV, a slepton mass less than 
4 TeV (3 TeV) can be probed at the $2\sigma$ ($5\sigma$) level, which is much beyond the LHC reach.

\section*{Acknowledgements}
This work was supported by the National Natural Science Foundation of China (NNSFC) under 
grant Nos.12075300, 11821505 and 11851303, 
by Peng-Huan-Wu Theoretical Physics Innovation Center (12047503), 
by the CAS Center for Excellence in Particle Physics (CCEPP), 
by the CAS Key Research Program of Frontier Sciences 
and from a Key R$\&$D Program of Ministry of Science and Technology under number 2017YFA0402204. 
CH acknowledges support from the Sun Yat-Sen University Science Foundation.


\begin{thebibliography}{99}

\bibitem{Aad:2020nyj}
G.~Aad \textit{et al.} [ATLAS],
JHEP \textbf{10} (2020), 062
doi:10.1007/JHEP10(2020)062
[arXiv:2008.06032 [hep-ex]].

\bibitem{Aad:2020sgw}
G.~Aad \textit{et al.} [ATLAS],
Eur. Phys. J. C \textbf{80} (2020) no.8, 737
doi:10.1140/epjc/s10052-020-8102-8
[arXiv:2004.14060 [hep-ex]].

\bibitem{Aad:2019vvi}
G.~Aad \textit{et al.} [ATLAS],
Phys. Rev. D \textbf{101} (2020) no.7, 072001
doi:10.1103/PhysRevD.101.072001
[arXiv:1912.08479 [hep-ex]].

\bibitem{Aad:2019qnd}
G.~Aad \textit{et al.} [ATLAS],
Phys. Rev. D \textbf{101} (2020) no.5, 052005
doi:10.1103/PhysRevD.101.052005
[arXiv:1911.12606 [hep-ex]].

\bibitem{LEP}
http://lepsusy.web.cern.ch/lepsusy/


\bibitem{CEPCStudyGroup:2018ghi}
J.~B.~Guimar\~aes da Costa \textit{et al.} [CEPC Study Group],
[arXiv:1811.10545 [hep-ex]].

\bibitem{ILC}
T.Behnkeetal.,arXiv:1306.6327[physics.acc-ph];
H.Baeretal.,arXiv:1306.6352[hep-ph]; 
C. Adolphsen et al., arXiv:1306.6353 [physics.acc-ph]; 
C. Adolphsen et al., arXiv:1306.6328 [physics.acc-ph]; 
T. Behnke et al., arXiv:1306.6329 [physics.ins-det].

\bibitem{Abada:2019zxq}
A.~Abada \textit{et al.} [FCC],
Eur. Phys. J. ST \textbf{228} (2019) no.2, 261-623
doi:10.1140/epjst/e2019-900045-4

\bibitem{Badziak:2017the}
M.~Badziak, M.~Olechowski and P.~Szczerbiak,
Phys. Lett. B \textbf{770} (2017), 226-235
doi:10.1016/j.physletb.2017.04.059
[arXiv:1701.05869 [hep-ph]].

\bibitem{Han:2014sya}
C.~Han,
Int. J. Mod. Phys. A \textbf{32} (2017) no.33, 1745003
doi:10.1142/S0217751X17450038
[arXiv:1409.7000 [hep-ph]].

\bibitem{Abdughani:2017dqs}
M.~Abdughani, L.~Wu and J.~M.~Yang,
Eur. Phys. J. C \textbf{78} (2018) no.1, 4
doi:10.1140/epjc/s10052-017-5485-2
[arXiv:1705.09164 [hep-ph]].



\bibitem{Kim:1983dt}
J.~E.~Kim and H.~P.~Nilles,
Phys. Lett. B \textbf{138} (1984), 150-154
doi:10.1016/0370-2693(84)91890-2

\bibitem{Barenboim:2014kka}
G.~Barenboim, E.~J.~Chun, S.~Jung and W.~I.~Park,
Phys. Rev. D \textbf{90} (2014) no.3, 035020
doi:10.1103/PhysRevD.90.035020
[arXiv:1407.1218 [hep-ph]].

\bibitem{ATLASCollaboration:2016wlb}
M.~Aaboud \textit{et al.} [ATLAS],
Eur. Phys. J. C \textbf{76} (2016) no.9, 517
doi:10.1140/epjc/s10052-016-4344-x
[arXiv:1606.09150 [hep-ex]].

\bibitem{Alwall:2011uj} 
  J.~Alwall, M.~Herquet, F.~Maltoni, O.~Mattelaer and T.~Stelzer,
  ``MadGraph 5 : Going Beyond,''
  JHEP {\bf 1106}, 128 (2011)
  doi:10.1007/JHEP06(2011)128
  [arXiv:1106.0522 [hep-ph]].

\bibitem{Sjostrand:2014zea} 
  T.~Sjöstrand {\it et al.},
  ``An Introduction to PYTHIA 8.2,''
  Comput.\ Phys.\ Commun.\  {\bf 191}, 159 (2015)
  doi:10.1016/j.cpc.2015.01.024
  [arXiv:1410.3012 [hep-ph]].

\bibitem{Shen:2019yhf}
Y.~Shen, H.~Xiao, H.~Li, S.~Qin, Z.~Wang, C.~Wang, D.~Zhang and M.~Ruan,
``Photon Reconstruction Performance at the CEPC baseline detector,''
[arXiv:1908.09062 [physics.ins-det]].

\bibitem{Gu:2017del}
J.~Gu and Y.~Y.~Li,
Chin. Phys. C \textbf{42}, no.3, 033102 (2018)
doi:10.1088/1674-1137/42/3/033102
[arXiv:1709.08645 [hep-ph]].

\bibitem{Burns:2008va}
M.~Burns, K.~Kong, K.~T.~Matchev and M.~Park,
JHEP \textbf{03}, 143 (2009)
doi:10.1088/1126-6708/2009/03/143
[arXiv:0810.5576 [hep-ph]].

\bibitem{Cho:2007qv}
W.~S.~Cho, K.~Choi, Y.~G.~Kim and C.~B.~Park,
Phys. Rev. Lett. \textbf{100}, 171801 (2008)
doi:10.1103/PhysRevLett.100.171801
[arXiv:0709.0288 [hep-ph]].

\bibitem{Matchev:2009ad}
K.~T.~Matchev and M.~Park,
Phys. Rev. Lett. \textbf{107}, 061801 (2011)
doi:10.1103/PhysRevLett.107.061801
[arXiv:0910.1584 [hep-ph]].

\bibitem{Konar:2009wn}
P.~Konar, K.~Kong, K.~T.~Matchev and M.~Park,
Phys. Rev. Lett. \textbf{105}, 051802 (2010)
doi:10.1103/PhysRevLett.105.051802
[arXiv:0910.3679 [hep-ph]].

\end{thebibliography}
\end{document}